\documentstyle[12pt,aaspp4]{article}
\def\kms{km s${}^{-1}$}
\def\ab{$\sim$}

\def\p{$\pm$}
\def\etal {{et~al.}}
\def\eg {{\it e.g.}}
\def\deg {$^\circ$}
\def\solmass{M$_\odot$}
\def\HI {H\kern0.1em{\sc i}}

\lefthead{Peck \& Taylor}
\righthead{A Circumnuclear Disk in 1946+708}
\slugcomment{to appear in {\it Astrophysical Journal Letters}}

\begin{document}

\title{Evidence for a Circumnuclear Disk in 1946+708}
\author{A. B. Peck\altaffilmark{1} and G. B. Taylor\altaffilmark{2}}
\altaffiltext{1} {Max-Planck-Institut f\"ur Radioastronomie, Auf dem H\"ugel 69, D-53121 Bonn, Germany, apeck@mpifr-bonn.mpg.de}
\altaffiltext{2}{National Radio Astronomy Observatory, P.O. Box O,
Socorro, NM 87801;\\gtaylor@nrao.edu}

\setcounter{footnote}{0}

\begin{abstract}
A number of extragalactic radio sources which exhibit symmetric jets
on parsec scales have now been found to have neutral hydrogen
absorption at or near the systemic velocities of their host galaxies.
Understanding the spatial distribution and kinematics of the \HI\
detected toward the central parsecs of these sources provides an
important test of unified schemes for AGN.
 
We present results of Global VLBI Network observations of the
redshifted 21 cm \HI\ line toward the Compact Symmetric Object
1946+708 ($z$=0.101).  We find significant structure in the gas on
parsec scales.  The peak column density of the \HI\ (N$_{\rm
HI}\sim$3$\times$10$^{23}$ cm$^{-2}$(${T_s}\over{\rm 8000K}$) ) occurs near the center of
activity of the source, as does the highest velocity dispersion
(FWHM$\simeq$ 350 to 400 km s$^{-1}$).  There is also good evidence
for a torus of ionized gas with column density 7 $\times$ 10$^{22}$
cm$^{-2}$.  The jets in 1946+708 exhibit bi-directional motion
measurable on timescales of a few years.  The resulting unique
information about the geometry of the continuum source greatly assists
in the interpretation of the gas distribution, which is strongly
suggestive of a circumnuclear torus of neutral atomic and ionized
material with one or more additional compact clumps of gas along the
line of sight to the approaching jet.

\end{abstract}
\keywords{galaxies:active -- galaxies:individual (1946+708) -- radio
continuum:galaxies -- radio lines:galaxies}

\section{Introduction}
Compact Symmetric Sources (CSOs) are a recently identified class of
radio sources which are less than 1 kpc in size, and are thought to be
very young objects ($\le$10$^4$~yr, Readhead \etal\ 1996; Owsianik \&
Conway 1998).  CSOs exhibit milliarcsecond scale jets marked by steep
spectrum hotspots (Taylor \etal\ 1996).  The jets are oriented at
large angles to the line of sight, resulting in very little Doppler
boosting of the approaching jet (Wilkinson \etal\ 1994).  This means
that the receding jet may be responsible for up to half of the
observed radio continuum flux density, and thus an obscuring torus,
such as that predicted by unified schemes of AGN, (see \eg\ Antonucci
1993), should be detectable in absorption toward the core and one or
both hotspots.

This class of sources is unique among radio galaxies in that
$\sim$50\% of CSOs searched thus far have detectable \HI\ absorption
($\tau\ge$0.01; Vermeulen 2001, Peck \etal\ 2000) toward the core.  We
present extremely high spatial resolution observations of the \HI\
absorption in the CSO 1946+708.  The radio axis in this source
is thought to be oriented between 65 and 80\deg\ to the line of sight,
based on VLBI radio continuum studies of the proper motions in the jet
components (Taylor \& Vermeulen 1997).  This source exhibits an ``S''
shaped symmetry, shown by the contours in Fig.~1.  The location
of the radio core is indicated with an asterisk.  The northern hotspot (NHS)
is located in the approaching jet, while the end of the receding jet
is marked by a southern hotspot (SHS).

\section{Observations and Analysis}

The observations were made on 1999 23 February using 13 antennas of
the Global VLBI Network (8 stations of the NRAO\footnote{The National Radio Astronomy Observatory is a facility of the National
Science Foundation operated under a cooperative agreement by
Associated Universities, Inc.} Very Long Baseline
Array (VLBA), the 27 element NRAO Very Large Array (VLA) in
phased-array mode, the NRAO Greenbank 140 ft telescope, the Effelsberg
100m telescope, the Westerbork tied-array, and the Onsala 25m
telescope) with 2 bit sampling.  A single IF with a bandwidth of 8 MHz
was observed in 256 channels, resulting in a frequency resolution of
31.25 kHz.  
Following bandpass
correction performed using 3C345 and 2344+8226, Doppler correction was
applied using the AIPS task CVEL.  At this point it was found that the
central frequency of the line does not match that shown in the VLBA
observations of 1995 (Peck, Taylor and Conway 1999, hereafter referred
to as PTC99) due to an error in the application of CVEL in the earlier
data set.  The central velocity is actually 220 \kms\ higher than that
indicated in PTC99.  The correct velocity is closer to the optical
systemic velocity of the host galaxy, and thus still consistent with
the model presented in PTC99.


\section{Results}

\subsection{Radio Continuum}

The radio structure of 1946+708 has been previously reported on by
Taylor \& Vermeulen (1997), and by PTC99.  The angular resolution of
Fig.~1 is 4.9 $\times$ 4.3 mas, or about 50\% better than that
obtained by PTC99 at the same frequency.  At $\sim$5 mas resolution
the core component, located midway between the hot spots, is still
blended with the inner jet components (identified as N5 and S5 in
Taylor and Vermeulen 1997) and the smooth underlying jet.  The
brightest feature in Fig.~1 is the NHS, but it is only a few times
brighter than the SHS or the string of jet components in between.
Faint and symmetric extensions are seen to emerge from the hotspots,
but they fade rapidly into the noise.

We have obtained observations of 1946+708 at 5 GHz from the
Caltech-Jodrell Bank survey program (CJF -- Taylor \etal\ 1996)
session on February 8, 1998.  The 5 GHz data were tapered and then
restored with the same beam as the 1.29 GHz image in order to produce
the spectral index image shown in Fig.~2.  This image is comparable
with that produced by PTC99, but our enhanced resolution reveals a
dramatic spectral turnover close to the center of activity.  In fact
the most strongly inverted region is about a fourth of a beam
southwest of the core component.  This region runs perpendicular to
the radio jet axis.  Since the jet has a deconvolved width of just 0.5
mas at 5 GHz, it is difficult to assess the significance of features
transverse to the jet, but we suspect that there is sufficient SNR at
the edges of the jet ($>$ 4:1) to support the claim that the region of
the low frequency spectral turnover extends across the jet.

\subsection{\HI~ Absorption}

Neutral hydrogen absorption is present toward all of the radio
components in the Compact Symmetric Object 1946+708.  The profiles
shown in Fig.~1 are integrated over 3$\times$3 pixels, an area
slightly smaller than the synthesized beam, toward the peak of each
continuum component identified in Taylor \& Vermeulen (1997).  The
smooth line in each panel indicates a multicomponent Gaussian fit to
the data.  In panels 1 through 4, two Gaussian components were used,
while in panels 5 \& 6, only one was used.  The parameters derived
from the Gaussian fits to the profiles are shown in Table 1.  The
optical depth has been estimated using T$_{\rm s}$=8000 K, as
described in PTC99.  The systemic velocity obtained from optical
observations of both emission and stellar absorption lines (Stickel \&
K\"{u}hr 1993) is 30279\p300 \kms.  Thus all of the \HI\ absorption
features reported here are at the systemic velocity to within the
uncertainty of the optical measurements.

A broad component is seen in all 6 profiles.  This broad line varies
significantly in width across the source, with the broadest region
appearing near the center of activity of the source (profile 4). The
narrowest region in the broad line occurs toward the continuum peak
seen in profile 2.  Although the line appears broad toward the
northeastern extremity of the jet in profile 1, is seems likely that
this is due to lower signal to noise in this region.  This profile
lacks the bluer wing seen in the other broad profiles, suggesting
that, although the gap between the broad and narrow line is not
apparent, the profile is more similar to that seen in profile 2 than
to profiles 3 through 6.  The opacity does not vary much, but
increases gradually toward the receding jet, while the column density
peaks near the core.  

A deep narrow line is distinguishable from the broad component in
profiles 1 through 4.  This line appears to be present toward profiles
5 \& 6 as well, but the higher optical depth of the broad line in this
region makes it impossible to fit with a separate Gaussian component.
The line has a FWHM linewidth of only 35$-$70 \kms\ and does not vary
much in central velocity.  These fits are shown as the (b) components
in the first four profiles in Table 1.

In addition to fitting Gaussian functions to the integrated \HI\
absorption profiles shown above, fits have also been made at each
pixel across the source.  The top row of plots in Fig.~3 contains the
parameters obtained from fits to the broader line.  Pixels with a peak
signal to noise ratio less than 2 have been blanked.  The broad
absorption occurs toward the entire continuum source, as seen in the
optical depth ($\tau$) distribution shown in the first panel.  The
second panel shows the velocity field, with centroids ranging from
30190 to 30340 \kms.  The linewidth distribution shown in the third
panel indicates a dramatic increase in velocity dispersion just to the
northeast of the core, where the FWHM is $>$300 \kms.  The second row
of plots shows the narrower line, which is only distinguishable from
the broad line over part of the source.  In this component, little
variation in $\tau$, velocity centroid or FWHM is seen.
\section{Discussion}

\subsection{Evidence for an atomic torus}

The \HI\ absorption in 1946+708 consists of a very broad line and a
lower velocity narrow line which are visible toward the entire \ab100
pc of the continuum source.  The broad line has low optical depth and
peaks in column density near the core of the source.  This is
consistent with a thick torus scenario in which gas closer to the
central engine is much hotter, both in terms of kinetic temperature
and spin temperature, so a longer path-length through the torus toward
the core would not necessarily result in a higher optical depth.  The
gas toward the core, however, would be rotating much more quickly
around the central massive object, and thus would have a higher
velocity dispersion and column density.  Given the known orientation
of the radio axis, close to the plane of the sky with the northern jet
on the approaching side, we would expect the broadest linewidths to
appear in projection just south of the core of the source.  The offset
to the north shown in Fig.~3 (panel 3) might indicate that the inner
part of the atomic torus is inclined by 20-30\deg\ from perpendicular
to the radio axis.  A schematic of this scenario is shown in Fig.~4.
Lu (1990) has suggested that an ``S'' shaped symmetry in radio jets
might be due to precession caused by an offset accretion disk, making
this scenario consistent with the observed morphology of the radio
continuum source.

The narrow line has a higher optical depth, and does not vary much
across the source.  The most likely explanation for this narrow line
is that it arises from gas further out in the torus, perhaps an
extended region of higher density gas on the order of at least 80-100
pc in diameter.  This cloud would have a higher optical depth than
much of the torus, occurring in a region where the shielding from the
central x-ray source is greater and thus T$_{\rm s}$ lower, and it
would also have a narrower linewidth, given that we would be seeing a
much smaller section of the annulus of rotating material at that
radius.  The FWHM of much of the gas in the narrow line is still
significantly greater than that seen toward the center of our own
Galaxy (\ab20 \kms; van der Hulst \etal\ 1983) and so it seems
unlikely that it is simply a cloud in the host galaxy which happens to
lie on the line of sight to the central radio source.  The slightly
higher velocity centroid of the broad line might be due to an inward
streaming motion of the gas in this region of the torus, or might be
indicative of a contribution from gravitational redshift if the gas is
within a few pc of a central object loosely estimated to have a mass
of 5$\times$10$^8$\solmass\ (PTC99).

So far, no molecular gas has been detected in this
source. Observations at the frequency of the redshifted OH line have
yielded an upper limit of $\tau\sim$0.006 (PTC99).

\subsection{Evidence for a torus of ionized gas}

The low frequency spectral turnover near the core presented in \S3
(see Fig.~2) could be the result of either synchrotron self-absorption
(SSA) of the emitting components, or free-free absorption by a
foreground disk of ionized gas.  In the case of SSA we expect the
frequency at which the turnover occurs to be related to the angular
size and flux density of the components.  Based on a flux density of
250 mJy, size of 0.8 mas measured at 5 GHz, and assuming equipartition
(Scott \& Readhead 1977), we derive a spectral turnover for the NHS at
an observed frequency of $\le$ 1.3 GHz.  This is roughly consistent
with the spectrum presented in PTC99, although the peak frequency is
poorly constrained since the spectrum of the NHS has not been measured
below 1.3 GHz.  For the weaker jet components of similar angular size,
the peak frequency should be well below 1 GHz.  For the inner jet
components this poses a problem for the SSA model, and requires
significant departures from equipartition conditions.

A more likely explanation for the spectral turnover is that it is
produced by free-free absorption from the ionized innermost region of
the circumnuclear torus.  This explanation was also explored by PTC99,
as shown in Fig.~8 of that paper.  The new observations reveal that
the opacity of the ionized gas reaches 2.2 toward the core.  This is
larger than the estimate of PTC99 of 0.4 owing to the higher
resolution afforded by the new observations which reduce blending of
the absorbed region with the unabsorbed steep spectrum jet on either
side.  The assumptions used in PTC99 (temperature $\sim$ 8000 K and a
path length of 50 pc) now imply a column density of 6.9 $\times$
10$^{22}$ cm$^{-2}$ and a density of 460 cm$^{-3}$ for the ionized material. 
The peak in the column density of ionized gas is not spatially
coincident with the peak velocity dispersion in the \HI\ gas,
indicating that the longest pathlength through the ionized region is
not precisely toward the innermost radius of the atomic torus, as
shown in Fig.~4.  This is consistent with model outlined in PTC99, and
with the expected ionization fraction of only 1-10\% in the atomic
torus (Neufeld \& Maloney 1995).

A similar case for a disk of free-free absorbing material has been
made for the CSO 0108+388 (Marr, Taylor and Crawford 2000).  A problem
for both the free-free absorption and the SSA models is that the
turnover in 1946+708 appears so broad (Fig.~2 inset).  This is
probably the result of blending of components with different spectra
within the synthesized beam, or a non-uniform opacity in the torus.

\section{Conclusion}

The high velocity dispersion toward the core of 1946+708 is indicative
of fast moving circumnuclear gas, perhaps in a rotating toroidal
structure.  Further evidence for this region of high kinetic energy
and column density is found in the spectral index distribution which
indicates a region of free-free absorption along the line of sight
toward the core and inner receding jet.  The \HI\ optical depth
increases gradually toward the receding jet.  The most likely scenario
to explain these phenomena consists of an ionized region around the
central engine, as well as an accretion disk or torus, with a scale
height of $\le$10 pc at the inner radius and at least 80 pc at the
outer radius, which is comprised primarily of atomic gas.

\begin{acknowledgements}

  We thank K. Menten and an anonymous referee for a critical reading
of the manuscript and F. Bertoldi for the flux density measurement
made at 250 GHz (shown in Fig.~2 inset) using the IRAM 30m telescope
at Pico Veleta.  This research has made use of the NASA/IPAC
Extragalactic Database (NED) which is operated by the Jet Propulsion
Laboratory, California Institute of Technology, under contract with
the National Aeronautics and Space Administration.

\end{acknowledgements}

\clearpage


\begin{figure}
\vspace{8cm}
\includegraphics{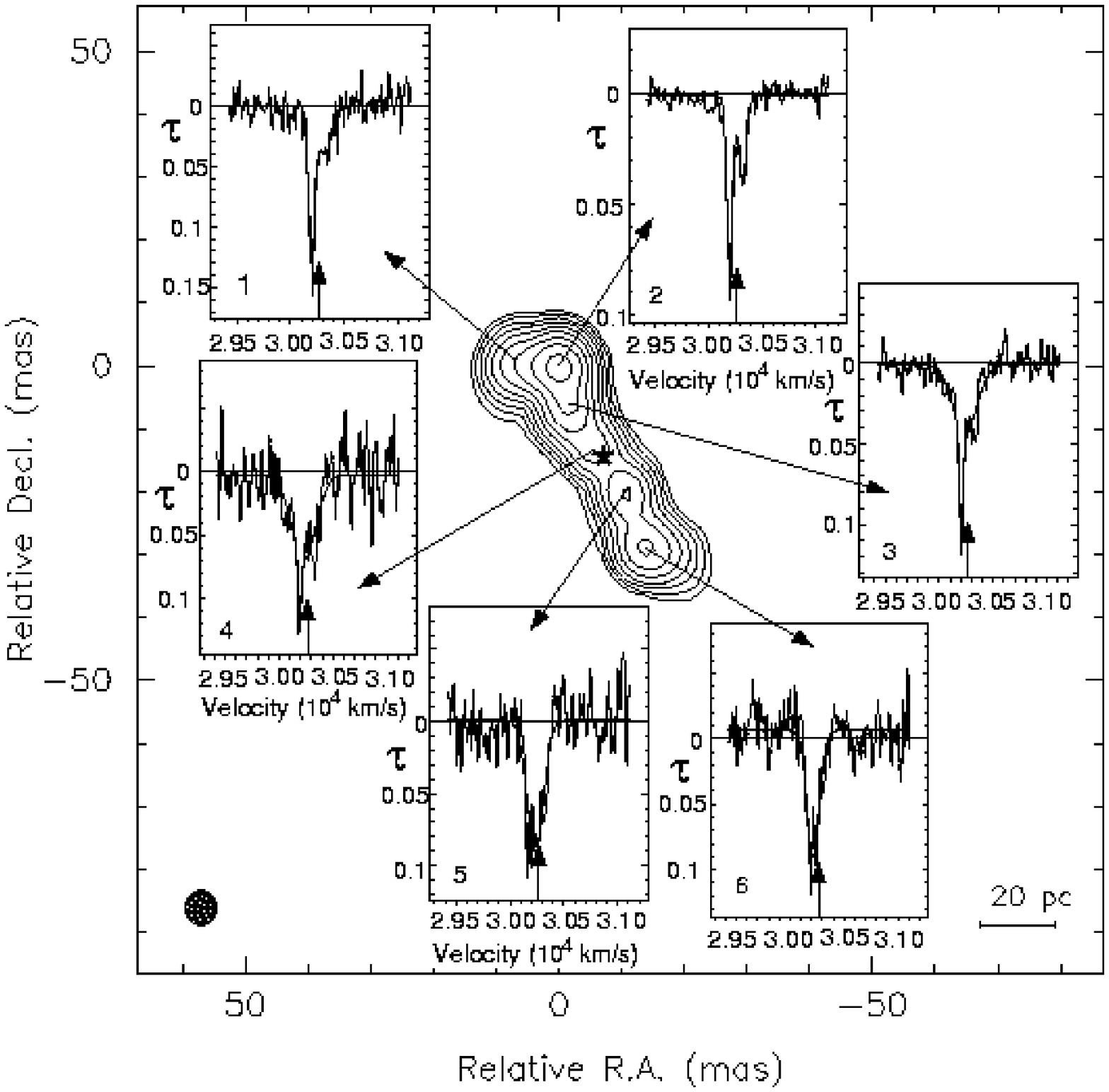}
\figcaption{The absorption profiles toward each of the 6 resolved
continuum components across the source.  The systemic velocity is
indicated by an arrow in each profile.  The velocity resolution is 16
\kms.  The rms noise is 0.8 mJy/beam/channel.  The 1.290 GHz continuum
contours shown begin at 1.5 mJy/beam and increase by factors of 2.
The peak is 429 mJy.  The position of the radio core is indicated with
an asterisk. The beam, shown in the lower left corner, is
4.3$\times$4.9 mas.  The linear scale shown in the lower right assumes
H$_0$=75 \kms\ Mpc$^{-1}$.
\label{fig1}}
\end{figure}

\begin{figure}
\vspace{8cm} 
\includegraphics{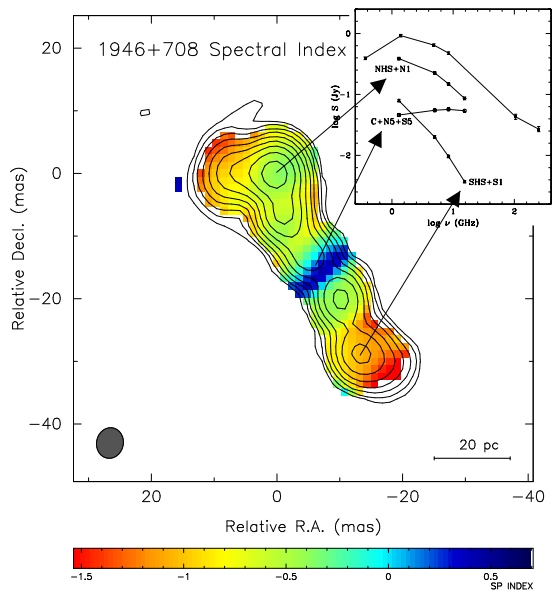} 
\figcaption{An image of the spectral index
distribution between 1.3 and 5 GHz. {\it Inset}: Continuum spectra
from single dish measurements of 1946+708 between 330 MHz and 250 GHz.
Also shown are groups of components, (labeled as in Taylor \&
Vermeulen 1997), measured between 1.3 and 15 GHz with the VLBA.
\label{fig2}}
\end{figure}

\begin{figure}
\vspace{18cm} 
\includegraphics{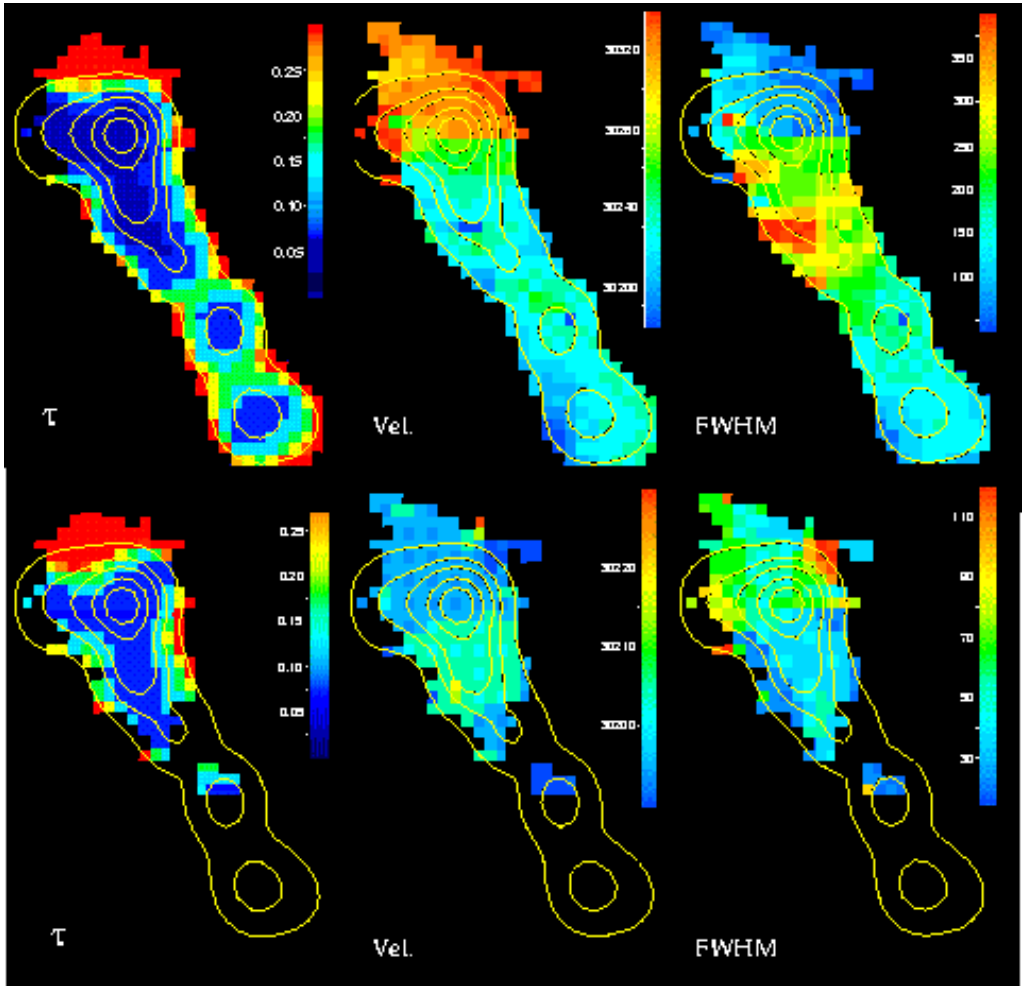} 
\figcaption{The results of Gaussian functions
fitted at each pixel where the signal to noise ratio (SNR) is greater
than 2.  {\bf Row 1:} Panel 1 shows the optical depth of the broad
line.  Values range from $\tau$=0.01 to 0.3, though the highest
values are likely due to poor fits in the regions of lowest SNR.
Panel 2 shows the velocity field of the broad line.  Velocity
centroids range from 30190 to 30340 \kms.  Panel 3 shows the FWHM of
the broad line.  Widths range from 80 to 375 \kms.  {\bf Row 2:} Panel
4 shows the optical depth of the narrow line.  Values range from
$\tau$=0.01 to 0.3, though again the highest values occur in the
regions of lowest SNR.  Panel 5 shows the velocity field of the narrow
line.  Velocity centroids range from 30180 to 30210 \kms.  Panel 6
shows the FWHM of the narrow line.  Widths range from 20 to 115 \kms.
\label{fig3}}
\end{figure}

\begin{figure}
\vspace{8cm} 
\includegraphics{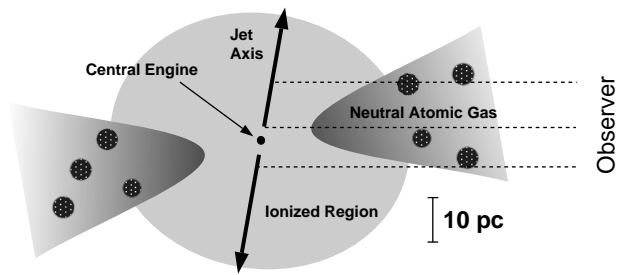} 
\figcaption{A cartoon showing the probable
orientation of the radio jets and circumnuclear torus in 1946+708.
The radio jet is thought to be oriented between 65 and 80\deg\ to our
line of sight.   With the atomic torus offset by $\sim$25\deg\ with
respect to the perpendicular, our line of sight through the inner part
of the torus intersects the radio jet slightly to the north of the
core, while the region of longest pathlength through the ionized gas
is toward the inner southern jet.  Our line of sight to the northern
(approaching) jet passes through part of the torus and also appears to
intersect one or more clumps of denser \HI.  The scale shown is
approximate, the inner radius of the torus and the sizes of the denser
clumps are not known.
\label{fig4}}
\end{figure}

\clearpage

\begin{table}
\begin{center}
\begin{tabular}{ccrcccc}
& & Amplitude & Central Velocity & FWHM & & N$_{\rm HI}$\tablenotemark{a} \\
Profile &Component& (mJy)~ & (\kms) & (\kms) & $\tau$ & (10$^{22}$ cm$^{-2}$) \\
\hline
1&a & 3.9\p0.4 & 30315\p15 & 166\p33 & 0.04\p0.011 &9.7 \\
&b & 10.4\p0.7 & 30195\p1.9 & 69.0\p5.3 &0.12\p0.017 &12.0  \\
\hline
2&a & 11.5\p0.6 & 30318\p2.1 & 84.9\p5.4 & 0.04\p0.005& 5.0 \\
&b & 23.8\p0.6 & 30198\p0.9& 63.1\p2.2 & 0.08\p0.008 & 7.4 \\
\hline
3&a & 5.9\p0.4 & 30237\p5.8 & 256\p15 & 0.04\p0.005 &14.9  \\
&b & 10.5\p0.7 & 30206\p1.2 & 35.9\p3.1 & 0.07\p0.008 &3.7 \\
\hline
4&a & 3.2\p0.3 & 30232\p9.9 & 264\p25 & 0.07\p0.007 &29.6  \\
&b & 3.0\p0.7 & 30192\p3.6 & 34.5\p9.5 &0.06\p0.026 &3.0 \\
\hline
5&a & 5.5\p0.3 & 30234\p5.0 & 181\p12 & 0.09\p0.010 &23.8 \\
\hline
6&a & 5.4\p0.3 & 30217\p3.5 & 136\p8.7 & 0.10\p0.011 &19.8 \\

\hline
\end{tabular}
\tablenotetext{a}{Assuming a spin temperature of 8000 K and a covering
factor of 1.}
\end{center}
\tablenum{1}
\caption{Gaussian Functions fitted to Absorption Profiles in
Each Region}
\end{table}

\end{document}